\documentclass[prd,superscriptaddress,amsfonts,amssymb,amsmath,showpacs,onecolumn]{revtex4-2}
\usepackage{bm}
\usepackage{amsfonts}
\usepackage{graphicx}
\usepackage{amsmath}
\usepackage{palatino}
\usepackage{mathpazo}
\usepackage{textcomp}
\usepackage[utf8]{inputenc}
\usepackage[T1]{fontenc}
\usepackage{float}
\usepackage{booktabs}
\usepackage{dcolumn}
\usepackage{multirow} 
\usepackage{hyperref}
\hypersetup{colorlinks,citecolor=blue}
\usepackage{amsmath}
\usepackage{xcolor}
\usepackage{orcidlink}
\usepackage[caption=false]{subfig}
\usepackage{commath}
\usepackage{amssymb}

\usepackage{esdiff}
\usepackage{mathtools}
\usepackage{ifthen}
\usepackage{mathtools, amssymb} 
\usepackage{siunitx}
\usepackage{tikz}
\usetikzlibrary{hobby} 
\usetikzlibrary{arrows.meta} 
\tikzset{>={Latex[length=4,width=4]}} 
\usetikzlibrary{calc,intersections,decorations.markings}
\usepackage{siunitx}
\usepackage{xcolor} 

\colorlet{mylightblue}{blue!20}
\colorlet{myblue}{blue!50!black}
\colorlet{mydarkblue}{blue!30!black}
\colorlet{mylightred}{red!10}
\colorlet{myred}{red!50!black}
\colorlet{mydarkred}{red!60!black}
\colorlet{mydarkgreen}{green!30!black}

\tikzset{
  midarr/.style={decoration={markings,mark=at position #1 with {\arrow{stealth}}},postaction={decorate}},
  midarr/.default=0.5
}

\def\jnl@style{\it}
\def\aaref@jnl#1{{\jnl@style#1}}

\def\aaref@jnl#1{{\jnl@style#1}}

\def\aj{\aaref@jnl{AJ}}                   
\def\apj{\aaref@jnl{ApJ}}                 
\def\apjl{\aaref@jnl{ApJ}}                
\def\apjs{\aaref@jnl{ApJS}}               
\def\apss{\aaref@jnl{Ap\&SS}}             
\def\aap{\aaref@jnl{A\&A}}                
\def\aapr{\aaref@jnl{A\&A~Rev.}}          
\def\aaps{\aaref@jnl{A\&AS}}              
\def\mnras{\aaref@jnl{Mon.~Not.~Roy.~Astron.~Soc.}}             
\def\prd{\aaref@jnl{Phys.~Rev.~D}}        
\def\prc{\aaref@jnl{Phys.~Rev.~C}}  
\def\prl{\aaref@jnl{Phys.~Rev.~Lett.}}    
\def\qjras{\aaref@jnl{QJRAS}}             
\def\skytel{\aaref@jnl{S\&T}}             
\def\ssr{\aaref@jnl{Space~Sci.~Rev.}}     
\def\zap{\aaref@jnl{ZAp}}                 
\def\nat{\aaref@jnl{Nature}}              
\def\aplett{\aaref@jnl{Astrophys.~Lett.}} 
\def\apspr{\aaref@jnl{Astrophys.~Space~Phys.~Res.}} 
\def\physrep{\aaref@jnl{Phys.~Rep.}}      
\def\physscr{\aaref@jnl{Phys.~Scr}}       
\def\commat{\aaref@jnl{Comm.~Math.~Phys.}}              
\def\science{\aaref@jnl{Science}}               
\def\cqg{\aaref@jnl{Classical Quant.~Grav.}}            
\def\jpcs{\aaref@jnl{JPCS}}                                     
\def\ijmpd{\aaref@jnl{Int.~J.~Mod.~Phys.~D}}                    
\def\grg{\aaref@jnl{Gen.~Relat.~Gravit.}}               
\def\rpp{\aaref@jnl{Rep.~Prog.~Phys.}}          
\def\npa{\aaref@jnl{Nucl.~Phys.~A}}        
\def\lrr{\aaref@jnl{Living Rev.~Rel.}}                   
\def\jcap{\aaref@jnl{J.~Cosmology Astropart.~Phys.}}    
\def\rmp{\aaref@jnl{Rev.~Mod.~Phys.}}   
\def\epjc{\aaref@jnl{Eur.~Phys.~J.~C}}

\hypersetup{linkcolor=blue}

\allowdisplaybreaks[1]

\begin{document}

\color{black}       

\title{Buchdahl quark stars within $f(Q)$ theory}

\author{Oleksii Sokoliuk\orcidlink{0000-0003-4503-7272}}
\email{oleksii.sokoliuk@mao.kiev.ua}
\affiliation{Main Astronomical Observatory of the NAS of Ukraine (MAO NASU),\\
Kyiv, 03143, Ukraine}
\affiliation{Astronomical Observatory, Taras Shevchenko National University of Kyiv, \\
3 Observatorna
St., 04053 Kyiv, Ukraine}

\author{Sneha Pradhan\orcidlink{0000-0002-3223-4085}}
\email{snehapradhan2211@gmail.com}
\affiliation{Department of Mathematics, Birla Institute of Technology and
Science-Pilani,\\ Hyderabad Campus, Hyderabad-500078, India.}

\author{P.K. Sahoo\orcidlink{0000-0003-2130-8832}}
\email{pksahoo@hyderabad.bits-pilani.ac.in}
\affiliation{Department of Mathematics, Birla Institute of Technology and
Science-Pilani,\\ Hyderabad Campus, Hyderabad-500078, India.}

\author{Alexander Baransky\orcidlink{0000-0002-9808-1990}}
\email{abaransky@ukr.net}
\affiliation{Astronomical Observatory, Taras Shevchenko National University of Kyiv, \\
3 Observatorna
St., 04053 Kyiv, Ukraine}

\date{\today}
\begin{abstract}

In the present paper, authors study the strange stars with the MIT Bag EoS admitting Buchdahl symmetry (non-singular and physically viable metric potential) in the both linear $f(Q)=aQ+b$ and non-linear $f(Q)=Q+aQ^b$ forms of modified symmetric teleparallel gravitation. In order to obtain the correct form of Buchdahl metric coefficients, they matched interior strange star spacetime with the exterior Schwarzschild vacuum spacetime. As a strange star candidate, PSRJ1416-2230 has been used with mass $M=1.69M_\odot$ and radius $R=9.69R_\odot$. For either linear and non-linear $f(Q)$ gravities, they probed Null, Dominant and Strong Energy Conditions as well as the radial, tangential Equation of State (EoS), gradients of the perfect fluid energy-momentum tensor elements, Tolman-Oppenheimer-Volkoff equilibrium condition, relativistic adiabatic index and causality conditions, surface redshift. It was found that strange stars in the linear and non-linear $f(Q)$ gravity show physically viable behavior, respect energy and causality conditions, has EoS in the bounds $0\leq\omega\leq1$ and surface redshift does not exceed 2, as expected.

\textbf{Keywords:} Compact objects; modified gravity; non-metricity; MIT bag EoS 

\end{abstract}

\maketitle

\section{Introduction} \label{sec:1}

Within the last few decades, interest in the modified theories of gravitation has grown exponentially. It is generally known that with the presence of additional matter fields (such as inflaton), General Theory of Relativity (further - GR) could successfully describe the universe evolution starting from the cosmic dawn up to the late time acceleration of the universe. However, even considering that fact, GR fails to recreate recent cosmological observations and could not solve various cosmological problems, such as dark energy and dark matter problems, Hubble tension etc. Moreover, GR is non-renormalizable theory and diverge with the present higher-loop contributions. In that case, aforementioned modified theories of gravitation could help significantly.

Modified theories of gravitation are generally represented by the modified Lagrangian density (apart from the non-Lagrangian theories, MOND), that is consequently modified by the introduction of the additional geometrodynamical terms in the Einstein-Hilbert action integral. There are a lot of such modified theories present at the moment. The most successful ones (in the sense of the cosmological viability) are $f(R)$, $f(R,T)$, $f(\mathcal{T})$ and $f(Q)$ gravities respectively. $f(R)$ gravity is namely represented by the arbitrary function of the Ricci scalar curvature $R$ and was originally presented in \cite{Buchdahl:1983zz}, discussed in details in the pioneering works of \cite{RevModPhys.82.451,Faraoni:2008mf}. This theory is of special interest since it could provide a geometric mechanism for the description of inflation \cite{Starobinsky:1980te,Brooker:2016oqa,Huang:2013hsb} and solve the dark energy problem \cite{Nojiri:2017ncd,Capoziello2011}. On the other hand, there exist the other way to describe the gravitational interactions - via the so-called torsion and non-metricity. Gravities that arise from such terms are called the GR analogues and their modified versions could be regarded to as namely $f(\mathcal{T})$ and $f(Q)$ gravities. Both gravities could be obtained with the use of non-standard metric-affine connections, namely Weitzenb\"ock and metric incompatible connections that differ from the GR Levi-Cevita connection. In the present paper, we have chosen the case with the modified symmetric teleparallel gravitation (alternatively called $f(Q)$ gravity). 

There was a lot of investigations done within the field of $f(Q)$ theory of gravitation. For example, some of $f(Q)$ forms have been restricted in the sense of cosmological viability (with the use of such observational constraints as $H(z)$ datasets,  Baryon Acoustic Oscillations (BAO), quasars and Gamma Ray Bursts and CMB), and it was found that they could effectively describe the late time accelerated expansion as well as precisely predict the current value of Hubble parameter $H_0$ \cite{PhysRevD.100.104027}. Apart from the FLRW cosmology, a wide variety of compact objects of singular and non-singular nature were investigated in the $f(Q)$ gravity. For example, spherically symmetric and stationary black hole spacetimes \cite{DAmbrosio:2021zpm}, traversable wormholes \cite{Parsaei:2022wnu,Sokoliuk:2022efj,Banerjee:2021mqk,Sharma:2021egn} and star-like objects without singularity present \cite{Mandal:2021qhx,Lin:2021uqa}. In the following, we will briefly discuss such non-singular star-like objects that we are going to study in the present work.

In the last decades, non-singular compact objects become of special interest, since such stellar configurations could help us to constrain relativistic equation of state and even constrain modified theories of gravity via the energy conditions, matter causality and stability. Besides, stability could be probed either with Tolman-Oppenheimer-Volkoff equilibrium condition or with adiabatic index. Authors in this study are going to work with the special kind of compact stars, namely quark (or strange) stars admitting MIT bag Equation of State. It was reported that such exotic compact objects could exist, among the candidates for such stellar objects with high densities are millisecond pulsars SAX J1808.4-3658 and RXJ185635-3754, well-known X-ray pulsar Her X-1 and X-ray burster 4U 1820-30, pulsar PSR J1416-2230 \cite{Leahy:2007we,Panotopoulos:2019zxv}. We are going to inspect strange star PSR J1416-2230 with the use of linear and non-linear forms of $f(Q)$. 

Our article is organised as follows: in the first Section (\ref{sec:1}) we provide a brief introduction into the topic of modified theories of gravity, provide an examples of such theories and more detailed discussion on the chosen case of MOG, namely $f(Q)$ symmetric teleparallel gravity. Moreover, we discuss the non-singular compact strange stars. In the Section (\ref{sec:2}) we correspondingly present the formalism of the $f(Q)$ gravity and it's new geometrodynamical terms, affine connection and field equations applied to the spherically symmetric and static metric tensor. Besides, we as well assume that our strange star admit Buchdahl symmetry and match interior spacetime with exterior Schwarzschild vacuum spacetime in order to obtain the valid forms of Buchdahl metric coefficients. In the third Section (\ref{sec:3}) we introduce the MIT Bag EoS that we will be working with. Finally, we start our investigation with the linear model in the Section (\ref{sec:4}), where we probe energy conditions, equation of state for both radial and tangential pressures, gradients of the perfect fluid metric tensor components, Tolman-Oppenheimer-Volkoff (TOV) equilibrium condition, adiabatic index and matter causality, surface redshift. The same procedure was applied to the non-linear case in the Section (\ref{sec:5}). Concluding remarks on the whole work and it's key points are presented in the Section (\ref{sec:6}).

\section{$f(Q)$ gravitation formalism} \label{sec:2}

In the present study we assume that compact star lives on the differentiable Lorentzian manifold $\mathcal{M}$ that could be described properly by the metric tensor $g_{\mu\nu}$, it's determinant $g$ and affine connection $\Gamma$ (which is not metric compatible, for the sake of STEGR existence):
\begin{equation}
    g=g_{\mu\nu}dx^\mu\otimes dx^\nu,
\end{equation}
where $\Gamma^\alpha_{\beta}$ is the connection one form, that could be rewritten in terms of one forms of Levi-Cevita connection, deformation and contorsion tensor \cite{Ortin:2015hya}:
\begin{equation}
\Gamma^{\alpha}_{\beta}=w^{\alpha}_{\beta}+K^{\alpha}_{\beta}+L^{\alpha}_{\beta}.
\end{equation}
Equation above could be rewritten as follows:
\begin{equation}
\Gamma^{\alpha}_{\mu\nu}=\gamma^{\alpha}_{\mu\nu}+K^{\alpha}_{\mu\nu}+L^{\alpha}_{\mu\nu},
    \label{eq:2}
\end{equation}
where $\gamma$, $K$ and $L$ in the equation (\ref{eq:2}) are respectively Levi-Cevita metric-compatible affine connection, contorsion and deformation tensors. Since we assume $f(Q)$ gravity formalism, gravitation sector is fully described by the symmetric teleparallelism, which arise from the non-metricity one form and related tensor:
\begin{equation}
    Q^\alpha_{\beta}=\Gamma_{(ab)},\quad Q_{\alpha\mu\nu}=\nabla_\alpha g_{\mu\nu},
\end{equation}
where symmetric part of the tensor is defined as follows:
\begin{equation}
    F_{(\mu\nu)}=\frac{1}{2}\bigg(F_{\mu\nu}+F_{\nu\mu}\bigg).
\end{equation}
In the case when contorsion vanishes, we will be left with only deformation tensor terms:
\begin{equation}
    Q_{\alpha\mu\nu}=-L^{\beta}_{\alpha\mu}g_{\beta\nu}-L^{\beta}_{\alpha\nu}g_{\beta\mu},
\end{equation}
where deformation tensor is
\begin{equation}
    L^\alpha_{\mu\nu}=\frac{1}{2}Q^{\alpha}_{\mu\nu}-Q_{(\mu\nu)}^{\alpha}.
    \label{eq:77}
\end{equation}
Main quantity of gravitation sector within STEGR formalism is non-metricity scalar, constructed of non-metricity and the so-called superpotential
\begin{equation}
    Q=-P^{\alpha\mu\nu}Q_{\alpha\mu\nu},
\end{equation}
where superpotential has the following complex form:
\begin{equation}
    P^{\alpha}_{\mu\nu}=\frac{1}{4}\bigg[2Q^{\alpha}_{(\mu\nu)}-Q^{\alpha}_{\mu\nu}+Q^\alpha g_{\mu\nu}-\delta^{\alpha}_{(i}Q_{j)}-\overline{Q}^\alpha g_{\mu\nu}\bigg].
\end{equation}
Here, $Q^{\alpha}=Q^\nu_{\alpha\nu}$ and $\overline{Q}_\alpha=Q^\mu_{\alpha\mu}$ are two independent traces of the non-metricity tensor $Q_{\alpha\mu\nu}=\nabla_\alpha g_{\mu\nu}$. Finally, while we already presented all of the necessary term of STEGR formalism, we could proceed further and present the modified action integral for $f(Q)$ gravity (where $f(Q)$ is the arbitrary function of non-metricity scalar, that describes gravitational interactions in the theory) \cite{Xu2019}:
\begin{equation}
    \mathcal{S}[g,\Gamma,\Psi_i]=\int d^4x \sqrt{-g}f(Q)+\mathcal{S}_{\mathrm{M}}[g,\Gamma,\Psi_i].
    \label{eq:10}
\end{equation}
Here, $\mathcal{S}_{\mathrm{M}}[g,\Gamma,\Psi_i]$ denotes the action integral that defines the contribution of additional matter field $\Psi_i$ minimally and non-minimally coupled to gravity to the total Einstein-Hilbert action integral. If we will vary equation (\ref{eq:10}) with respect to the metric tensor inverse $g^{\mu\nu}$, we will get the abstract field equations for the theory:
\begin{equation}
\frac{2}{\sqrt{-g}}\nabla_\gamma\left(\sqrt{-g}\,f_Q\,P^\gamma\;_{\mu\nu}\right)+\frac{1}{2}g_{\mu\nu}f \\
+f_Q\left(P_{\mu\gamma i}\,Q_\nu\;^{\gamma i}-2\,Q_{\gamma i \mu}\,P^{\gamma i}\;_\nu\right)=-T_{\mu\nu},
\label{eq:11}
\end{equation}
where $f_Q\equiv\frac{df}{dQ}$ and $T_{\mu\nu}$ is usual energy momentum tensor, which general form could be written down as:
\begin{equation}
    T_{\mu\nu}=-\frac{2}{\sqrt{-g}}\frac{\delta(\sqrt{-g} \mathcal{L}_{\mathrm{M}})}{\delta g^{\mu\nu}}.
\end{equation}
In the equation above, $\mathcal{L}_{\mathrm{M}}$ denotes the Lagrangian density of matter fields such that $\int d^4x \sqrt{-g} \mathcal{L}_{\mathrm{M}}=\mathcal{S}_{\mathrm{M}}[g,\Gamma,\Psi_i]$. Also, by varying the action w.r.t. the affine connection $\Gamma^\alpha_{\,\,\,\,\mu\nu}$ we obtain:
\begin{equation}
\nabla_\mu \nabla_\nu \left(\sqrt{-g}\,f_Q\,P^\gamma\;_{\mu\nu}\right)=0.
\label{eq:12}
\end{equation}
In the following subsection we are going to derive the exact form of field equations (\ref{eq:11}) and (\ref{eq:12}) defined above for spherically symmetric objects.

\subsection{Field equations for spherically symmetric objects within $f(Q)$ theory}

The formula for field equation in $f(Q)$ gravity is\\
 \begin{equation}
 \kappa T_{tt} = \frac{e^{\nu-\lambda}}{2r^2}  [2rf_{QQ} Q'(e^{\lambda} -1)+ f_{Q}[(e^{\lambda} -1)(2+r\nu')+(1+e^{\lambda})r \lambda']+fr^{2}e^{\lambda}] ,
 \end{equation}
  \\
   \begin{equation}
 \kappa T_{rr} =- \frac{1}{2r^2}  [2rf_{QQ} Q'(e^{\lambda} -1)+ f_{Q}[(e^{\lambda} -1)(2+r\nu'+r\lambda')-2r\nu']+fr^{2}e^{\lambda}], \\
 \end{equation}
  \begin{equation}
 \kappa T_{\theta\theta} = - \frac{r}{4e^{\lambda}}  [-2rf_{QQ} Q' \nu' + f_{Q}[2\nu'(e^{\lambda} -2)-r\nu'^{2}+\lambda'(2e^{\lambda}+r\nu')-2r\nu'']+2fre^{\lambda}], \\
 \end{equation}
 where we have taken the spherically symmetric metric which is\\
 \begin{equation}
    ds^2= e^{\nu} dt^2-e^{\lambda} dr^2-r^2(d\theta^2+ sin^2 \theta d\phi^2).
    \end{equation}
The energy-momentum tensor for perfect fluid matter distribution is therefore looks exactly like
\begin{equation}
        T_{\mu\nu}=diag(e^{\nu}\rho, e^{\lambda}p_{r},r^{2}p_{t}, r^2 p_{t}sin^{2}\theta).
    \end{equation}
    Now by putting the value of components of energy-momentum tensor in the above field equations yields the result  \\
\begin{equation}
 8\pi\rho = \frac{1}{2r^2 e^{\lambda}}  [2rf_{QQ} Q'(e^{\lambda} -1)+ f_{Q}[(e^{\lambda} -1)(2+r\nu')+(1+e^{\lambda})r \lambda']+fr^{2}e^{\lambda}] ,
 \label{eq:18}
 \end{equation}
 \begin{equation}
 8\pi p_{r} = - \frac{1}{2r^2 e^{\lambda}}  [2rf_{QQ} Q'(e^{\lambda} -1)+ f_{Q}[(e^{\lambda} -1)(2+r\nu'+r\lambda')-2r\nu']+fr^{2}e^{\lambda}] ,
 \label{eq:19}
 \end{equation}
  \begin{equation}
 8\pi p_{t} = - \frac{1}{4re^{\lambda}}  [-2r f_{Q Q} Q' \nu' + f_{Q}[2\nu'(e^{\lambda} -2)-r\nu'^{2}+\lambda'(2e^{\lambda}+r\nu')-2r\nu'']+2f re^{\lambda}]. \\
 \end{equation}
Consequently, non-metricity scalar reads
\begin{equation}
    Q =\frac{1}{r}(\nu'+\lambda')(e^{-\lambda}-1).
\end{equation}
Now we could adopt the special form of metric potential $e^{\lambda(r)}$ to simplify the numerical calculations.

\subsection{Buchdahl metric}

\begin{figure}[!htbp]
    \centering
    \includegraphics[width=0.65\textwidth]{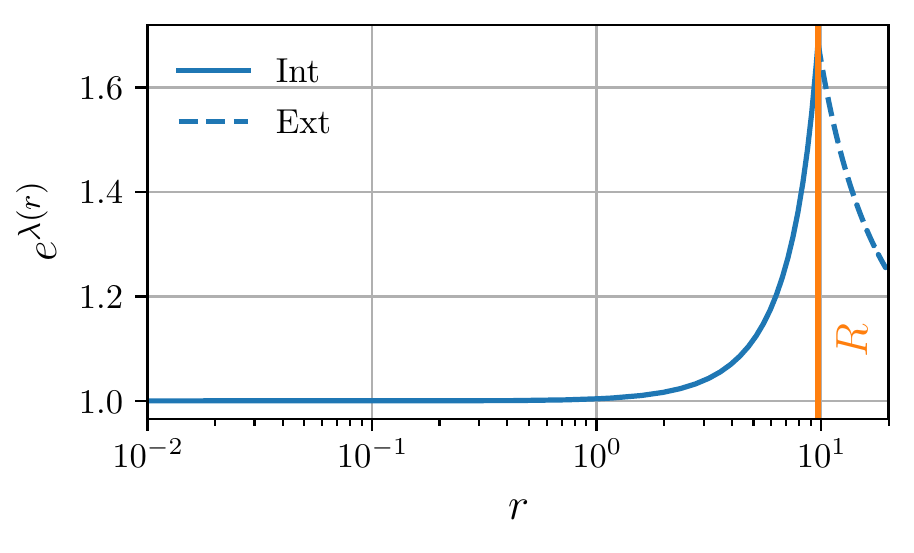}
    \caption{Buchdahl metric potential matched with Schwarzschild vacuum for $K=0.1$, $R=9.69\mathrm{km}$ and $M=1.97M_{\odot}$ (PSR J1416-2230)}
    \label{fig:1}
\end{figure}

As for the special form of metric potential $e^{\lambda(r)}$, we have chosen the case with Buchdahl metric potential, which looks exactly like \cite{KumarPrasad:2019rpk}:
\begin{equation}
    e^{\lambda(r)}=\frac{K(Cr^2+1)}{K+Cr^2},\quad 0<K<1 ,
\end{equation}
where $C$ is the metric function free parameter. This metric function and it's radial derivative is non-singular at the stellar origin, as required:
\begin{equation}
    e^{\lambda(0)}=1,\quad \partial_r e^{\lambda(r)}\big\rvert_{r=0}=0.
\end{equation}
Moreover, we could find the constant $C$ from the junction conditions (we are going to match the interior spacetime with exterior Schwarzschild vacuum spacetime at the stellar surface $\Sigma$, where $r=R$, $M$ is the total stellar mass enclosed within the sphere of radius $R$):
\begin{equation}
    \mathrm{Continuity\;of\;}g_{rr}:\quad \bigg(1-\frac{2M}{R}\bigg)^{-1}=\frac{K(Cr^2+1)}{K+Cr^2},
\end{equation}
\begin{equation}
    \mathrm{Continuity\;of\;}\frac{\partial g_{rr}}{\partial r}:\quad -\frac{2M}{(R-2M)^2}=\frac{2 C (K-1) K R}{\left(C R^2+K\right)^2}.
\end{equation}
Solution for the above equations is therefore given below:
\begin{equation}
    C=-\frac{2 K M}{R^2 (2 K M-K R+R)} .
\end{equation}
We could notice that for valid values of stellar mass and stellar radius, $C$ has negative values, as expected. Besides, we plot the matched metric potential for interior and exterior regions with $K=0.1$ on the Figure \ref{fig:1}. As one could easily notice, with fairly small and positive values of free parameter $K$ (as required by the model), we could obtain physically viable form of Buchdahl metric potential. On the aforementioned figure, for the sake of simplicity we marked the junction surface with orange vertical line and internal, external metric potentials are respectively plotted as solid and dashed lines.

In the further investigation, to solve the field equations defined above we need to assume the special Equation of State (EoS) for our matter distribution.

\section{MIT bag EoS and compact stars} \label{sec:3}

To simplify the calculation, one could assume special form of the so-called Equation of State. In the current paper, we will adopt the MIT bag EoS. This model is based on an assumption that the asymptotically free quarks are trapped in a region with finite volume, which is called bag (from the observational constraints, MIT bag constant has the bounds $41\mathrm{MeV/fm}^3<\mathcal{B}<75\mathrm{MeV/fm}^3$ \cite{1986ApJ...310..261A,doi:10.1146/annurev.ns.38.120188.001113}). The bag constant $\mathcal{B}$ exist as a inward pressure that traps the quarks inside the bag. In the MIT bag model, for the sake of simplicity it is assumed that the up, down and strange quarks are non-interacting ones and mass less. Thus, in the MIT bag model, radial pressure is given by:
\begin{equation}
    p_r = \sum_{f=u,d,s}p^f - \mathcal{B} ,
\end{equation}
and energy density consequently is:
\begin{equation}
    \rho = \sum_{f=u,d,s}\rho^f  + \mathcal{B} ,
\end{equation}
where the $p^f$ is the radial pressure for each flavor, $\rho^f$ is the energy density for each flavor. By combining the equations above, we could get the well-known simplified MIT bag EoS model:
\begin{equation}
    p_r = \frac{1}{3}(\rho - 4\mathcal{B}).
    \label{eq:24}
\end{equation}
We are going to assume that $\mathcal{B}=0.0001052\mathrm{km}^{-2}$, following the work of \cite{Chakraborty2021QuarkMS}.
Therefore, with the use of equation (\ref{eq:24}) and special form of the arbitrary function $f(Q)$ we could numerically analyse the behavior of compact stellar solutions.

\section{Linear model $f(Q)=aQ+b$}\label{sec:4}

\begin{figure}[!htbp]
    \centering
    \includegraphics[width=0.65\textwidth]{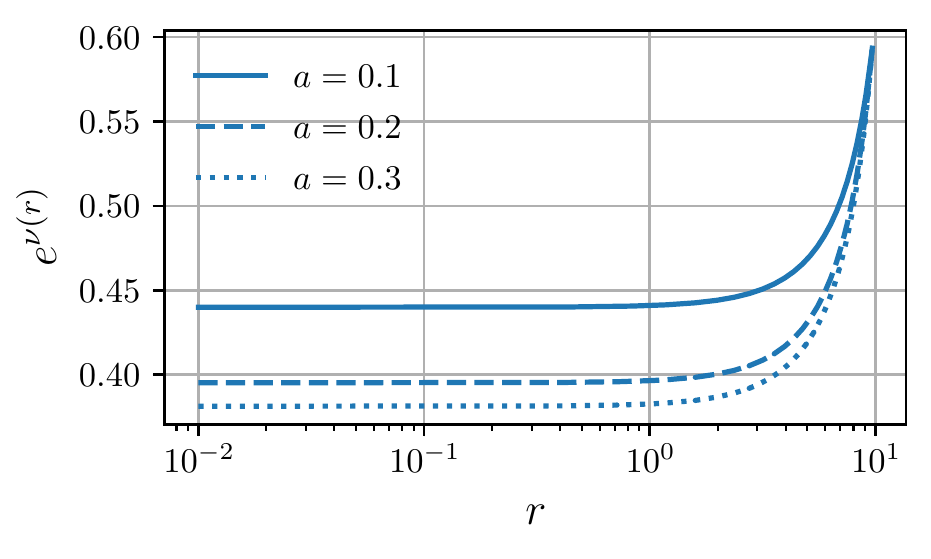}
    \caption{Numerically derived from field equation metric potential $e^{\nu(r)}$ with $K=0.1$ for linear $f(Q)$ gravity with vanishing $b=0$}
    \label{fig:2}
\end{figure}

The first model of $f(Q)$ theory that we are going to consider is the well-known linear gravitation:
\begin{equation}
    f(Q)=aQ+b,\quad f_Q=a,\quad f_{QQ}=0 ,
\end{equation}
where $a$ and $b$ are arbitrary constants such that $a\land b\in\mathbb{R}$, namely additional degrees of freedom. Correspondingly, in order to obtain the metric functions, we solve field equations (\ref{eq:18}) and (\ref{eq:19}) with initial condition $e^{\nu}\big\rvert_{r=R}=e^{-\lambda}\big\rvert_{r=R}$. Numerical solutions for metric potential is correspondingly placed on the Figure (\ref{fig:2}) with varying MOG parameter $a$ and vanishing $b$. As we see, this metric potential behaves as expected, converge with $e^{-\lambda}$ at the stellar surface, and therefore both metric potentials show physically viable behavior within the whole Lorentzian manifold.

\subsection{Energy conditions}

\begin{figure}[!htbp]
    \centering
    \includegraphics[width=\textwidth]{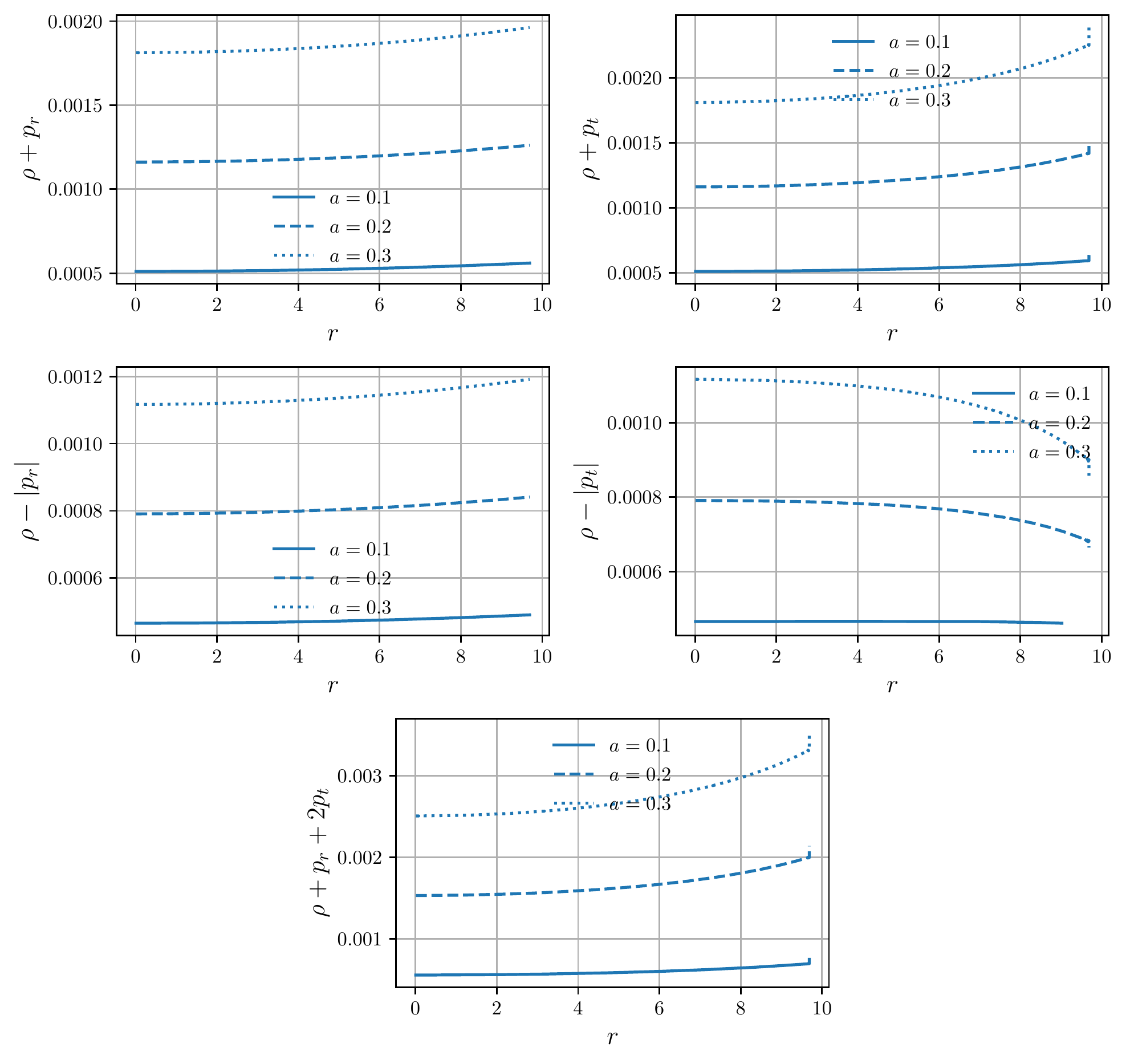}
    \caption{Null, Dominant and Strong energy conditions for linear $f(Q)$ gravity quark PSR J1416-2230 with $K=0.1$, vanishing $b=0$}
    \label{fig:3}
\end{figure}

Energy conditions are oftenly used to probe the viability of relativistic matter behavior. In general, there are four different kinds of them, namely Null Energy Condition (NEC), Weak Energy Condition (WEC), Strong Energy Condition (SEC) and Dominant Energy Condition (DEC). As it was noticed in the earlier works (see extensive discussion on the subject in \cite{Creminelli:2006xe,Hochberg:1998ha} and references therein), NEC arise as a minimal requirement of both WEC and SEC and needs to be validated everywhere if matter is considered to be viable. Aforementioned energy conditions could be easily written down in the terms of energy density and anisotropic pressure (those definitions could be obtained from the timelike and null-like Raychaudhuri equations, derivation of which could be found in the papers \citep{Rajabi:2021bdd,PhysRevD.86.083515}):
\begin{itemize}
\item Null Energy Condition (NEC): $\rho +p_{r}\geq 0$ and $\rho +p_{t}\geq 0
$

\item Weak Energy Condition (WEC) $\rho >0$ and $\rho +p_{r}\geq 0$ and $%
\rho +p_{t}\geq 0$

\item Dominant Energy Condition (DEC): $\rho -|p_{r}|\geq 0$ and $\rho
-|p_{t}|\geq 0$
\item Strong Energy Condition (SEC): $\rho +p_{r}+2p_{t}\geq 0$
\end{itemize}
Results of an investigation of energy conditions for quark star analogue of PSR J1416-2230, with $K=0.1$ and vanishing $b=0$ (linear $f(Q)$ gravity) are respectively plotted on the Figure \ref{fig:3}. As one could easily notice, it is obvious that each of the aforementioned energy conditions (NEC, DEC and SEC) are satisfied everywhere within stellar interior for every $a\geq0$ if $0<K<1$ (as required).  

\subsection{Equation of state and $\partial_r\rho$, $\partial_rp$}

\begin{figure}[!htbp]
    \centering
    \includegraphics[width=\textwidth]{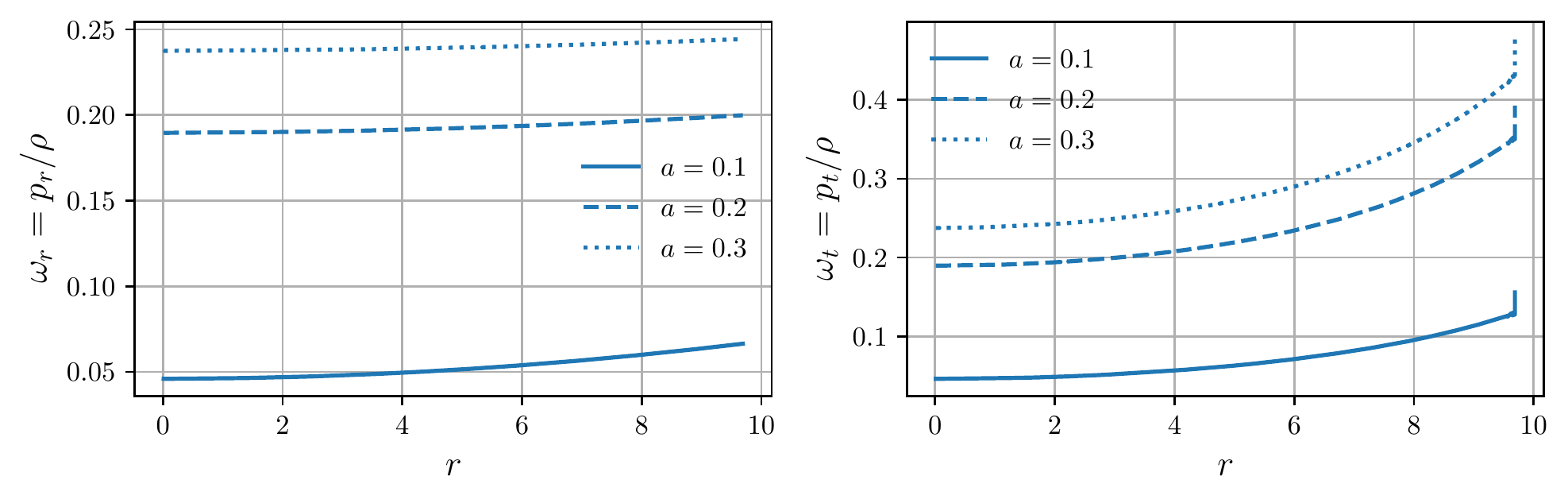}
    \caption{Anisotropic Equation of State for linear $f(Q)$ gravity quark PSR J1416-2230 with $K=0.1$, vanishing $b=0$}
    \label{fig:4}
\end{figure}

The next physical quantities that are we going to probe is the so-called equation of state parameter $\omega$ and radial gradients of the energy density and anisotropic pressure. Anisotropic EoS parameter is respectively given by those definitions:
\begin{equation}
    \omega_r=\frac{p_r}{\rho},\quad \omega_t=\frac{p_t}{\rho}
    \label{eq:31}.
\end{equation}
Solutions for above equations are plotted on the Figure \ref{fig:4}. As we noticed during the numerical analysis, both radial and tangential EoS behave like a viable cosmological fluid, i.e. $\omega$ values lie in the range $0\leq\omega\leq1$, (where vanishing EoS means dust and unitary means Zeldovich stiff fluid). 

\begin{figure}[!htbp]
    \centering
    \includegraphics[width=\textwidth]{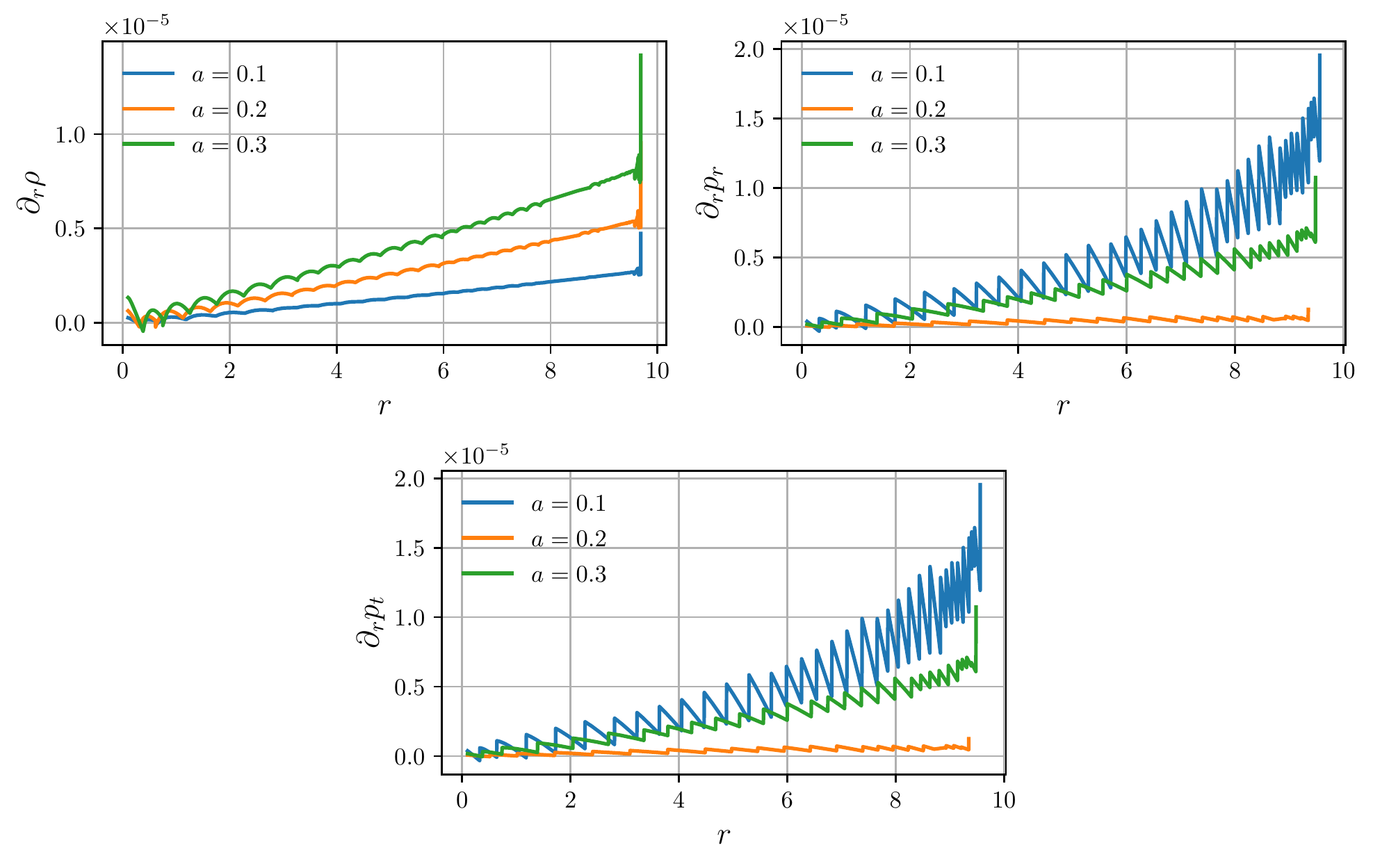}
    \caption{Gradients of energy-momentum tensor components for linear $f(Q)$ gravity quark PSR J1416-2230 with $K=0.1$, vanishing $b=0$}
    \label{fig:5}
\end{figure}

Moreover, we as well probe the behavior of energy-momentum tensor components radial gradients, that are properly plotted on the Figure \ref{fig:5}. It is obvious that gradients are positive and behave viably. In the next subsection, we are going to investigate the quark stars in terms of their stability.

\subsection{Tolman-Oppenheimer-Volkoff equilibrium}

\begin{figure}[!htbp]
    \centering
    \includegraphics[width=\textwidth]{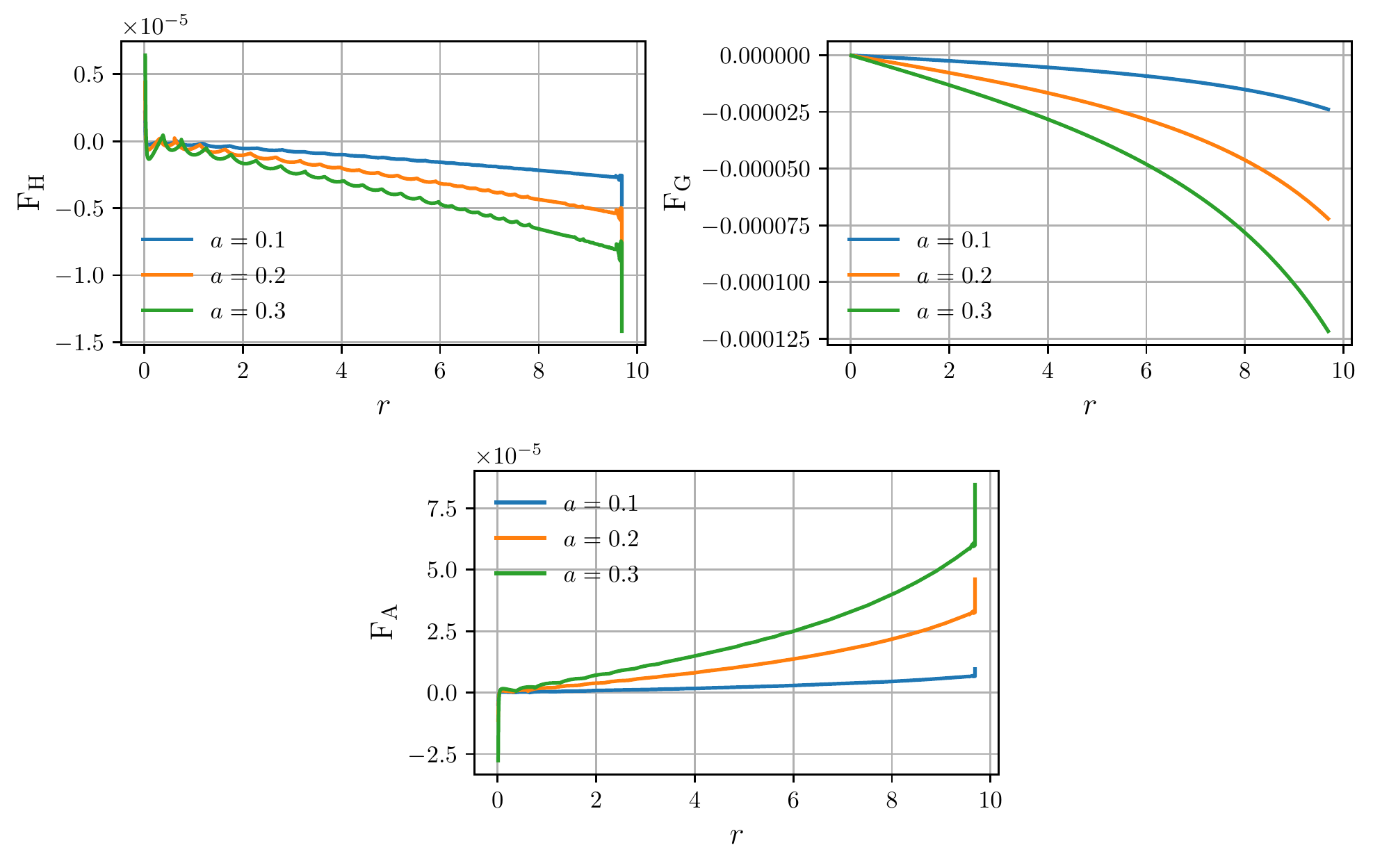}
    \caption{Forces present in TOV for linear $f(Q)$ gravity quark PSR J1416-2230 with $K=0.1$, vanishing $b=0$}
    \label{fig:6}
\end{figure}

Stability of the matter distribution within the stellar interior could be investigated throughout the well known Tolman-Oppenheimer-Volkoff equilibrium, which is given below in its modified form \cite{Oppenheimer1939,Poncede1993,Rahaman2014,Tolman1939}:
\begin{equation}
-\frac{dp_{r}}{dr}-\frac{\nu^{'}(r)}{2}(\rho+p_{r})+\frac{2}{r}(p_{t}-p_{r})+F_{\mathrm{ex}}=0
\label{eq:32}.
\end{equation}
Remarkably, equation above deviates from the standard TOV equilibrium by the presence of extra force, namely $F_{\mathrm{ex}}$, that arise because of the non-conserved stress-energy-momentum tensor ($\nabla^\mu T_{\mu\nu}\neq0$). Besides, we could separate equation (\ref{eq:32}) on the distinct forces, namely hydrodynamical, gravitational, anisotropical and extra forces:
\begin{equation}
F_H=-\frac{dp_{r}}{dr},\;\;\;\;\;\;\;\;F_A=\frac{2}{r}(p_{t}-p_{r}), \;\;\;\;\;\;\;\;F_G=-\frac{\nu'}{2}(\rho+p_{r}).
\label{eq:33}
\end{equation}
Numerical solutions for the aforementioned forces, defined in equation (\ref{eq:33}) are correspondingly placed on the Figure \ref{fig:6} for the different values of $a$. As we noticed, solutions are non-stable, but in the presence of arbitrary small extra force (with positive sign) they could be stable.

\subsection{Adiabatic index}

\begin{figure}[!htbp]
    \centering
    \includegraphics[width=0.65\textwidth]{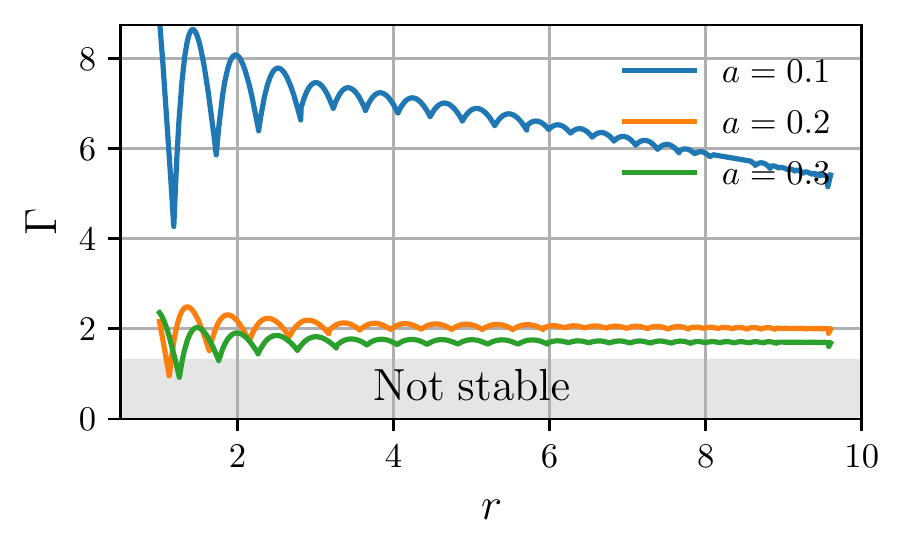}
    \caption{Anisotropic adiabatic index for linear $f(Q)$ gravity quark PSR J1416-2230 with $K=0.1$, vanishing $b=0$}
    \label{fig:7}
\end{figure}

Adiabatic index, derived in the pioneering work of Chandrasekhar \cite{PhysRevLett.12.114} is usually used to probe compact relativistic solutions in the sense of adiabatic perturbations stability. For anisotropic matter distribution, adiabatic index has the following straightforward form \cite{2017EPJC...77..328M}:
\begin{equation}
    \Gamma = \frac{p_r+\rho}{p_r}\frac{dp_r}{d\rho},
    \label{eq:34}
\end{equation}
where
\begin{equation}
    \frac{dp_r}{d\rho}=\frac{dp_r}{dr}\frac{dr}{d\rho}.
\end{equation}
Solution is considered to be adiabatically stable if inequality $\Gamma>4/3$ holds within the whole interior stellar region. Consequently, numerical solutions for adiabatic index are respectively represented on the Figure \ref{fig:7}. As we see, for relatively small and positive values of $a$, quark stars are adiabatically stable everywhere within the stellar interior. Besides, they are not stable near the stellar core for bigger $a$, and in the limit $a\to\infty$ they are not stable everywhere.

\subsection{Surface redshift}
Finally, to investigate the viability of the stellar model, one could also use the well-known surface redshift as a probe:
\begin{equation}
    \mathcal{Z}_s = |g_{tt}|^{-1/2}-1.
    \label{eq:35}
\end{equation}
For anisotropic matter, surface redshift must not exceed the value of 2. Consequently, we as usual numerically solve equation (\ref{eq:34}) and show it's solution on the first plot of Figure \ref{fig:8} for PSR J1416-2230. As we see, for positive values of $a$, surface redshift is positive and $\mathcal{Z}_s<2$. Therefore, this constraint is also satisfied for the model of our consideration.

\subsection{Quark matter causality}

\begin{figure}[!htbp]
    \centering
    \includegraphics[width=\textwidth]{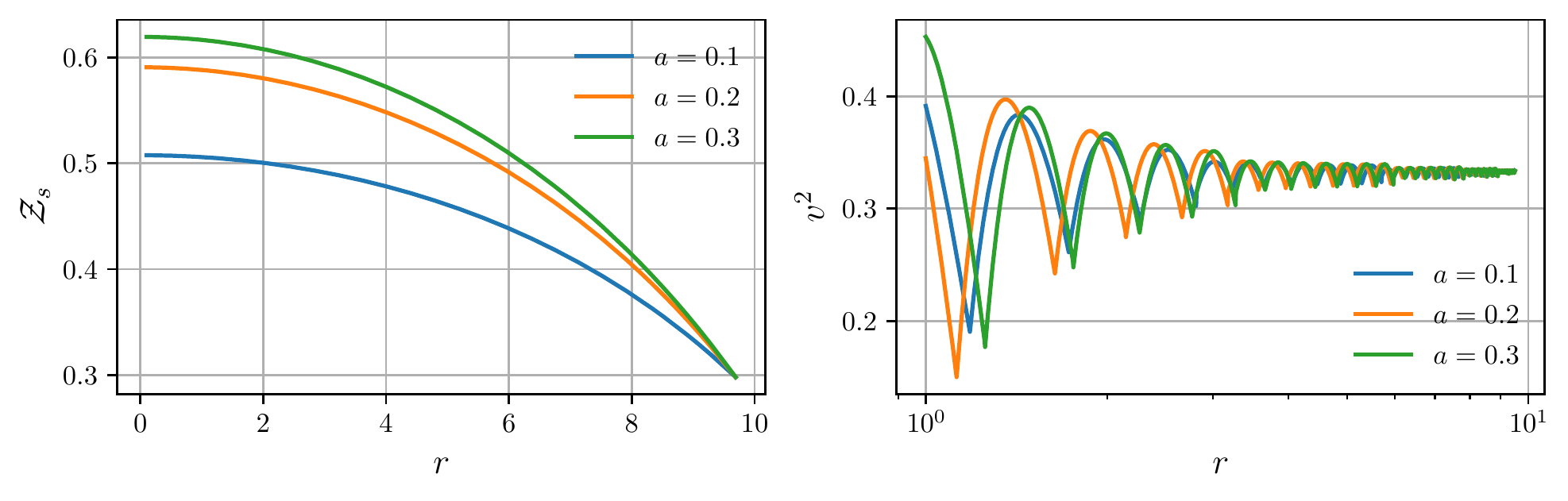}
    \caption{(\textit{left plot}) Surface redshift with $K=0.1$, vanishing $b=0$, (\textit{right plot}) Speed of sound squared for linear $f(Q)$ gravity quark PSR J1416-2230 with $K=0.1$, vanishing $b=0$}
    \label{fig:8}
\end{figure}

Causality of the matter could be probed via the speed of sound:
\begin{equation}
    v^2 = \frac{dp_r}{d\rho}\leq c^2 =1.
    \label{eq:37}
\end{equation}
We illustrate the squared speed of sound in the units of $c^2$ on the second plot of Figure \ref{fig:8}. As we see, quark matter do respect causality constraints. Therefore, solutions for linear $f(Q)$ gravity presented in the present and previous subsections are viable, both in the sense of energy conditions, stability and causality.

\section{Non-linear model $f(Q)=Q+aQ^b$}\label{sec:5}

\begin{figure}[!htbp]
    \centering
    \includegraphics[width=0.65\textwidth]{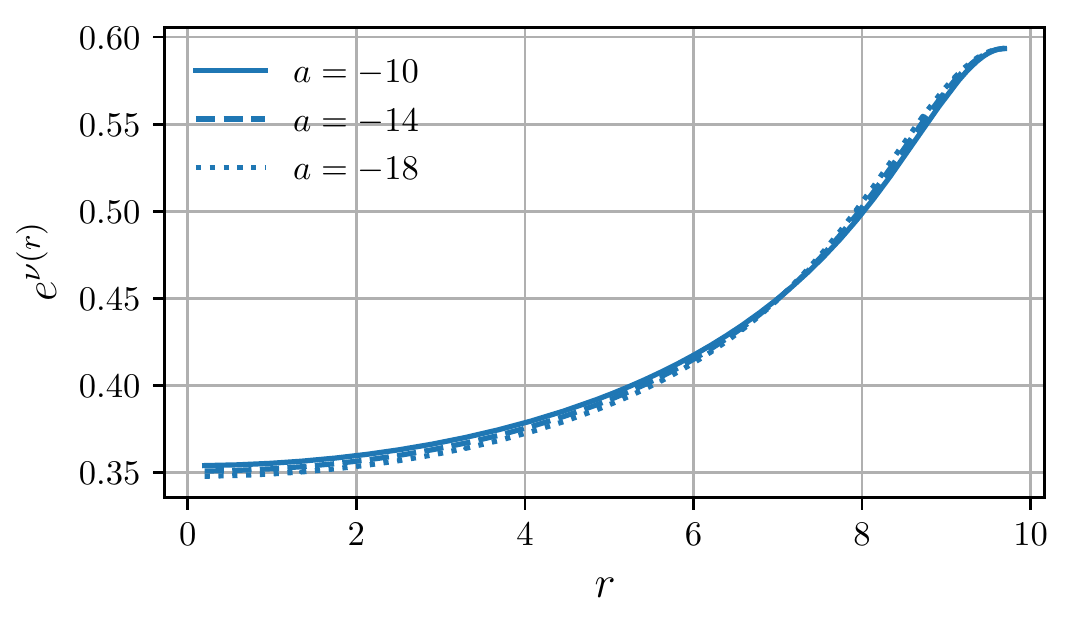}
    \caption{ Numerically derived from field equation metric potential $e^{\nu(r)}$ with $K=0.1$,non vanishing $b=2$ for non linear gravity}
    \label{fig:9}
\end{figure}

In this section we are considering the second model of $f(Q)$ theory, which is the non-linear one:\\
\begin{equation}
    f(Q)=Q+aQ^b,\quad f_Q=1 + a b Q^{(-1 + b)},\quad f_{QQ}=a (-1 + b) b Q^{(-2 + b)} ,
\end{equation}
where $a$ and $b$ are arbitrary constants such that $a\land b\in\mathbb{R}$, namely additional degrees of freedom.
Again in order to obtain the metric functions, we solve field equations (\ref{eq:18}) and (\ref{eq:19}) with initial condition $e^{\nu}\big\rvert_{r=R}=e^{-\lambda}\big\rvert_{r=R}$. Numerical solutions for metric potential is correspondingly placed on the Figure \ref{fig:9} with varying MOG parameter $a$ and non vanishing $b$. As we see, this metric potential behaves similar to the linear model one, converge with $e^{-\lambda}$ at the stellar surface.

\subsection{Energy conditions}

\begin{figure}[!htbp]
    \centering
    \includegraphics[width=\textwidth]{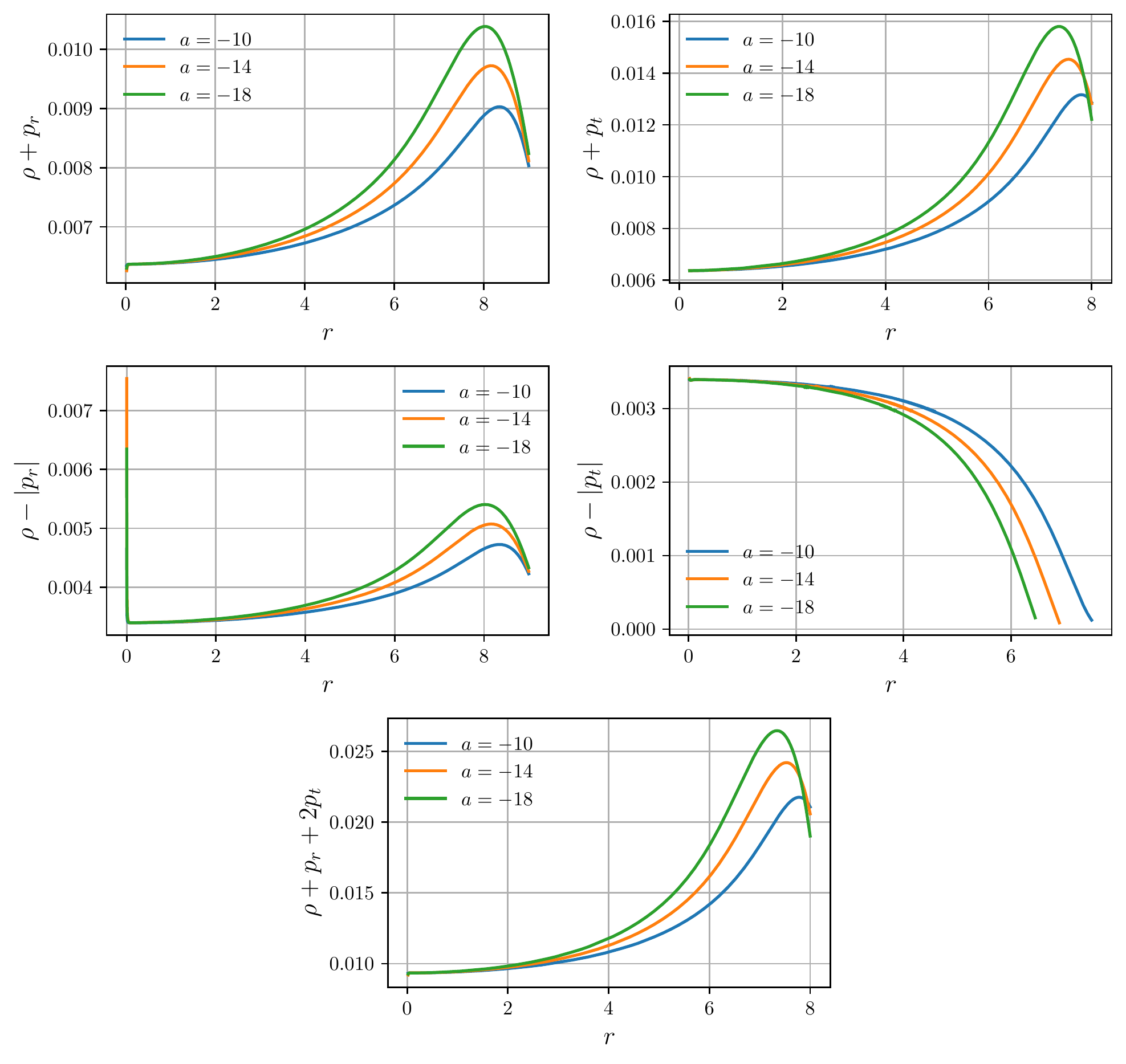},
    \caption{ Null, Dominant and Strong energy conditions for nonlinear $f(Q)$ gravity with $K=0.1$, non vanishing $b=2$}
    \label{fig:10}
\end{figure}

Now we plot the different energy conditions for Buchdahl quark star PSR J1416-2230 by taking the non linear model of $f(Q)$ gravity. As one can obviously notice from the Figure \ref{fig:10}, NEC is satisfied for both tangential and radial pressures. Moreover, DEC is validated for both pressure kinds, and finally SEC is also obeyed everywhere in the stellar region. So it is worth to say Null, Dominant and Strong Energy Conditions are fully satisfied for different values of negative a.

\subsection{Equation of state and $\partial_r\rho$, $\partial_rp$}

\begin{figure}[!htbp]
    \centering
    \includegraphics[width=\textwidth]{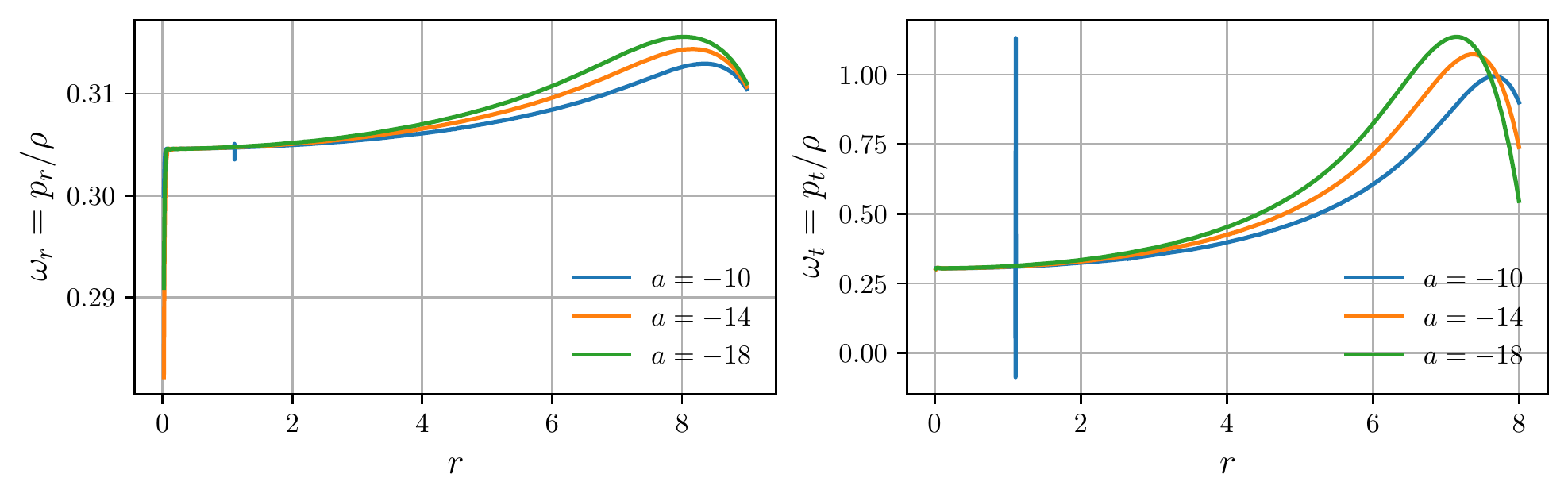}
    \caption{Anisotropic Equation of State for non linear $f(Q)$ gravity quark PSR J1416-2230 with $K=0.1$, non vanishing $b=2$}
    \label{fig:11}
\end{figure}

\begin{figure}[!htbp]
    \centering
    \includegraphics[width=\textwidth]{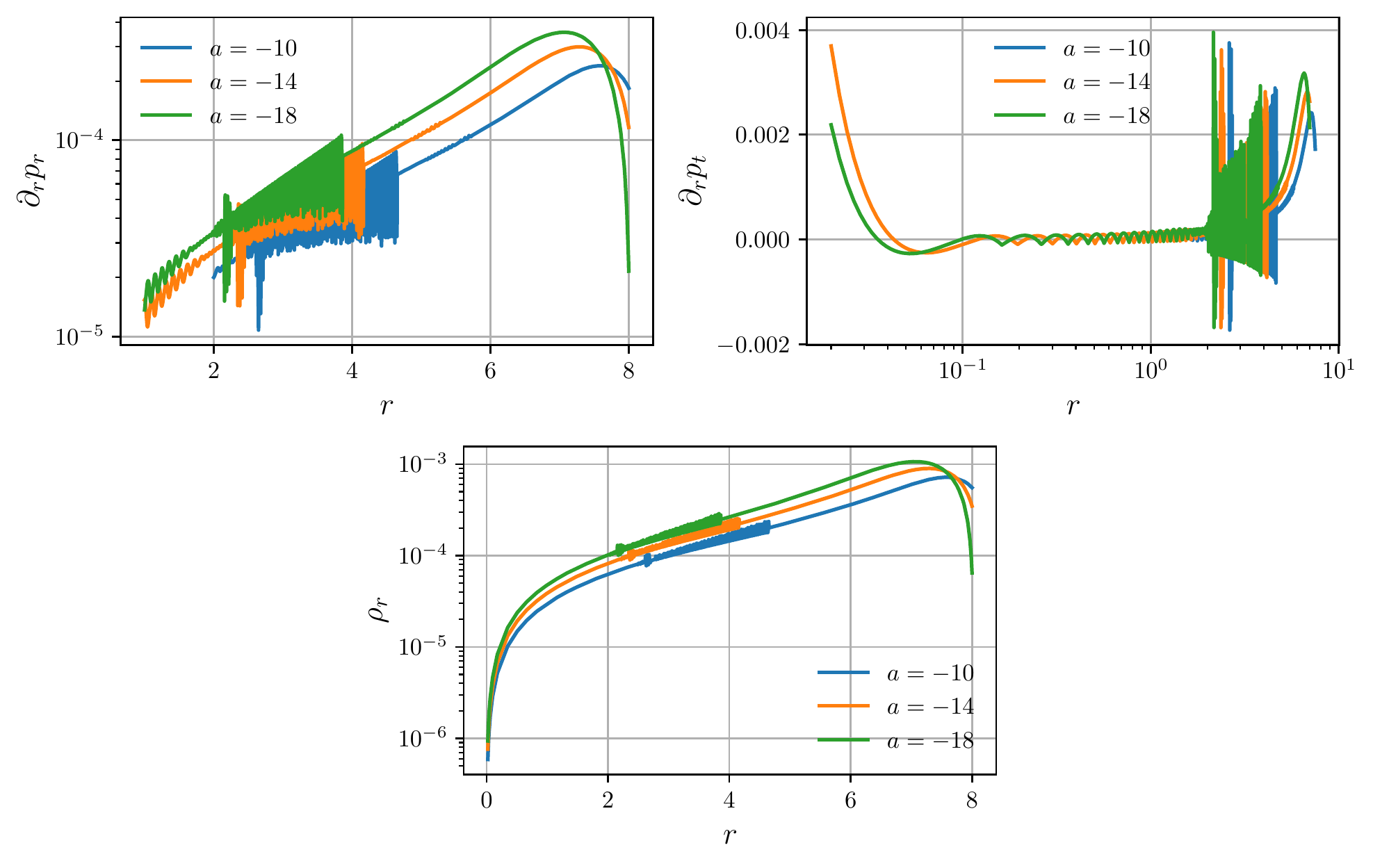}
    \caption{Gradients of energy-momentum tensor components for non linear $f(Q)$ gravity quark PSR J1416-2230 with $K=0.1$, non vanishing $b=2$}
    \label{fig:12}
\end{figure}

To examine the the behaviour of quark matter in the interior region of the Buchdahl quark star PSR J1416-2230 we are going to discuss the Equation of state parameters which were defined in the equation (\ref{eq:31}) earlier. As well, we could depict the energy density, anisotropic pressure gradients by simply evaluating $\rho'(r)$,  $p_{r}'(r)$, $p_{t}'(r)$. We illustrate EoS and stress-energy tensor component gradients on the Figures \ref{fig:11}, \ref{fig:12}. Here, we can see that the radial equation of state i.e. $\omega_{r}$ behaves as viable cosmological fluid as it lies in $0<\omega<1$. On the other hand, tangential equation of state parameter for non-linear model lies slightly outside of 1 for extremely small values of a.  

\subsection{Tolman-Oppenheimer-Volkoff equilibrium}

\begin{figure}[!htbp]
    \centering
    \includegraphics[width=\textwidth]{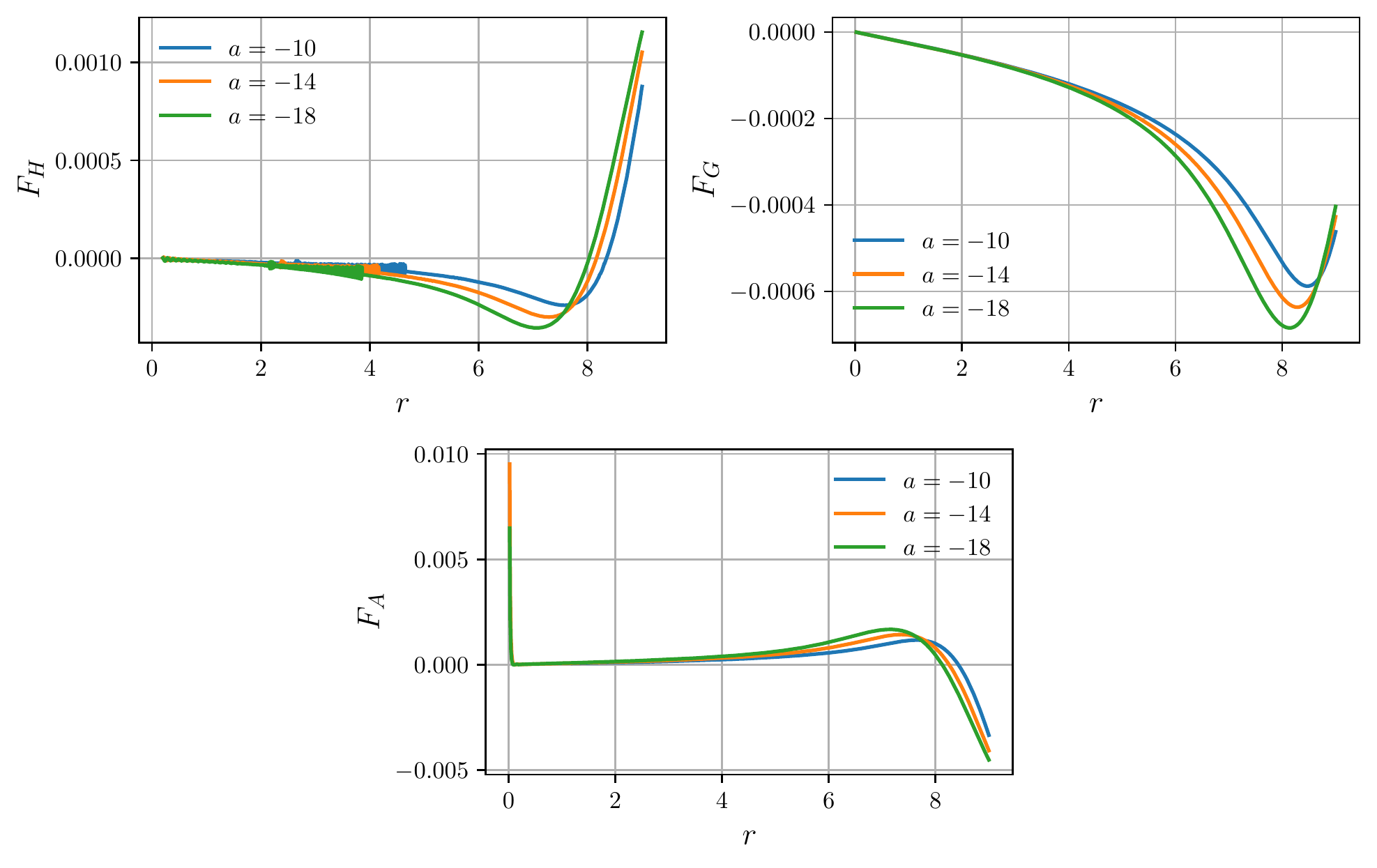}
    \caption{Forces present in TOV for non linear $f(Q)$ gravity quark PSR J1416-2230 with $K=0.1$, non vanishing $b=2$}
    \label{fig:13}
\end{figure}

Now the stability of the matter content in the stellar interior region can be investigated by using the well known Tolman-Oppenheimer-Volkoff equilibrium condition, which were given in the equation (\ref{eq:32}) in its modified form.  As well, in classical and modified TOV’s there present three additional forces: hydrodynamical $F_H$, gravitational $F_G$, and anisotropic $F_A$ which are mentioned in equation (\ref{eq:33}). So one can easily rewrite MTOV equation (\ref{eq:32}) as
\begin{equation}
    F_A+F_G+F_H+F_{ex}=0.
\end{equation}
We plot the solutions for different kind of forces with various values of $a$ on the Figure \ref{fig:13}. As we see, the hydrodynamical force $F_H$ is negative within the stellar core region. However, it is positive as it approaches the surface of the stellar region. The gravitational force $F_G$ is negative everywhere and the anisotropic force $F_A$ is positive from core to the intermediate stellar regions, but occurring negative nearby the stellar surface.

\subsection{Adiabatic index}

\begin{figure}[!htbp]
    \centering
    \includegraphics[width=0.65\textwidth]{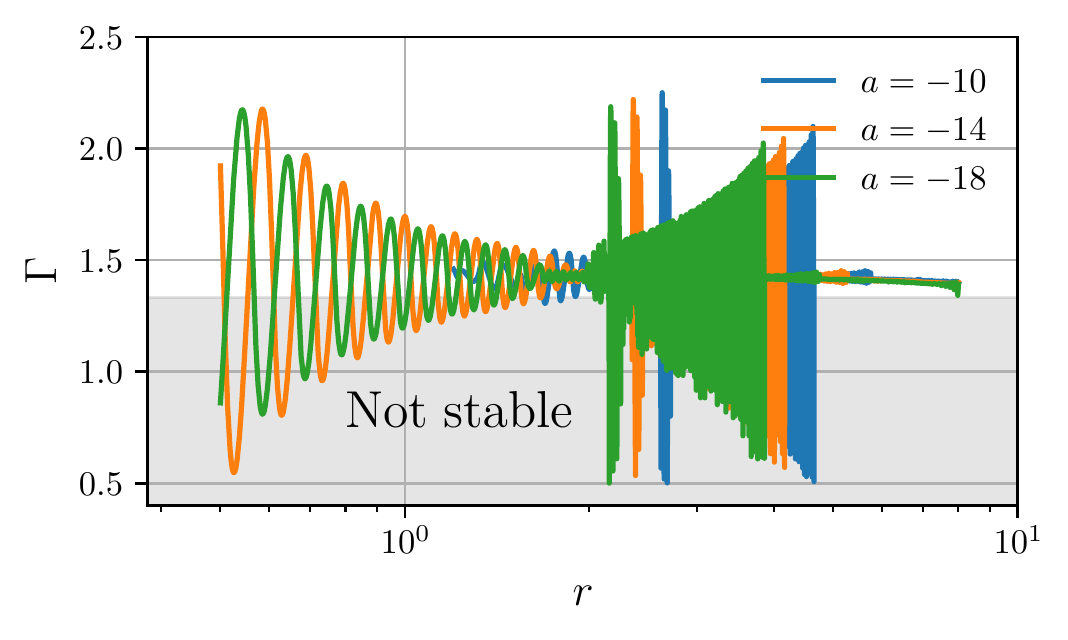}
    \caption{Anisotropic adiabatic index for non linear $f(Q)$ gravity quark PSR J1416-2230 with $K=0.1$, non vanishing $b=2$}
    \label{fig:14}
\end{figure}

Adiabatic index solutions are plotted on the Figure \ref{fig:14} for the special case of non-linear gravity. As we already mentioned, it is widely used to probe relativistic object via adiabatic perturbations. We notice that for any values of free parameter $a$ some regions of its is unstable ($\Gamma<4/3$) and some parts are stable ($\Gamma>4/3$). Besides, overall behavior of adiabatic index $\Gamma$ is very oscillatory because of the more complex form of function $f(Q)$.

\subsection{Surface redshift}

Finally, as earlier we define surface redshift as:
\begin{equation}
    \mathcal{Z}_s = |g_{tt}|^{-1/2}-1.
    \label{eq:34}
\end{equation}
According to the surface redshift constraints, we have plotted numerical solution for surface redshift for different values of $a$ on the first plot of Figure \ref{fig:15}. On the aforementioned plot one can see that surface redshift is positive and it does respect inequality $\mathcal{Z}_s<2$, besides it is gradually decreasing as it approaches the surface and finally converging at $r=R$.
 
\subsection{Quark matter causality}

\begin{figure}[!htbp]
    \centering
    \includegraphics[width=\textwidth]{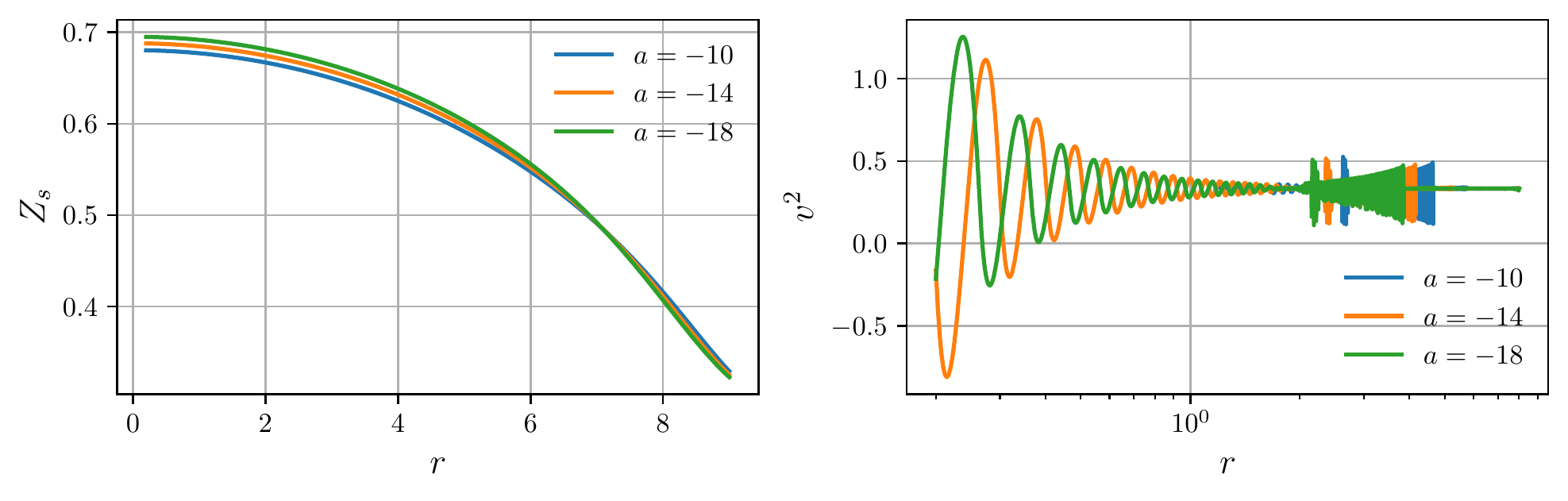}
    \caption{(\textit{left plot}) Surface redshift with $K=0.1$, vanishing $b=0$, (\textit{right plot}) Speed of sound squared for non linear $f(Q)$ gravity quark PSR J1416-2230 with $K=0.1$, non vanishing $b=2$}
    \label{fig:15}
\end{figure}

Another important criterion of matter content stability is the so-called speed of sound. This quantity is defined in equation  (\ref{eq:37}). Solution for speed of sound is plotted on the second plot of Figure \ref{fig:15}. As we see, the necessary condition that  the inequality $v^2\leq c^2=1$ always holds.
Therefore, we can conclude that solutions for non-linear $f(Q)$ gravity that are illustrated in the present and previous
subsections are stable both in the sense of energy conditions, causality. But in the case of Adiabatic index it is quite unstable as we have seen in Figure \ref{fig:14}.

\section{Concluding remarks}\label{sec:6}

In this paper, we comprehensively investigated the spherically symmetric and static compact stars that admit MIT Bag EoS within the context of $f(Q)$ gravity. We assumed two cosmologically viable cases of the $f(Q)$ functional form, namely linear $f(Q)=aQ+b$ and non-linear $f(Q)=Q+aQ^b$ gravities. In order to obtain the metric coefficients, matched with the external Schwarzschild vacuum spacetime, we used the strange star candidate PSRJ1416-2230 with mass $M=1.69M_\odot$ and radius $R=9.69R_\odot$. In the following, we are going to briefly present the key results of our investigation:
\begin{itemize}
    \item \textbf{Energy Conditions}: for linear form of $f(Q)$ each of the energy conditions that we consider (NEC, DEC and SEC) were validated everywhere within the stellar interior spacetime for every positive $a$ and vanishing $b$. On the other hand, for non-linear case, only NEC, SEC and radial WEC were obeyed everywhere for negative $a$ and $b=2$, tangential DEC was violated at the stellar envelope, but this could be solved in the limit $a\to0$. In order to study EC's in details, refer to the Figures \ref{fig:3} and \ref{fig:10}, corresponding subsections of the paper.
    \item \textbf{Equation of State}: the next quantity of our investigation is the EoS for both tangential and radial anisotropic pressures. It was found that for linear $f(Q)$ gravity, EoS of the strange star is physically viable (i.e. $0\leq\omega\leq1$) for each pressure kinds. However, for non-linear gravity (Starobinsky analogue in the modified symmetric teleparallel gravitation) only radial EoS is viable, tangential has $\omega_t>1$ near the envelope of the strange star. Graphical representation of the aforementioned results are located on the Figures \ref{fig:4} and \ref{fig:11}.
    \item \textbf{Gradients of the metric tensor components}: as well, we probed such quantities as $\partial_r \rho$, $\partial_r p_r$ and $\partial p_t$. For linear and non-linear $f(Q)$, as one could easily notice, gradients has monotonously growing tend up to the envelope and $\partial_r \rho\ll1$, $\partial_r p_r\ll1$, $\partial_r p_t\ll1$ which could ensure that speed of sound will not exceed the speed of light. Plots for the perfect fluid stress-energy-momentum component gradients are respectively located on the Figures \ref{fig:5} and \ref{fig:12} for varying values of free parameters.
    \item \textbf{TOV equilibrium}: in order to investigate the dynamical stability of compact stellar objects, we used the well-known TOV equilibrium condition. It was found that for strange star within the both linear and non-linear forms of $f(Q)$ to be stable, arbitrary small extra force needs to be present (alternatively, precisely fitted values of $a$ and non-vanishing $b$ could also lead to the strange star stability). As usual, results are properly illustrated on the Figures \ref{fig:6} and \ref{fig:13}.
    \item \textbf{Adiabatic index}: Adiabatic index has been widely used to constrain compact objects, and probe their stability from continuous adiabatic perturbations. It appears that linear $f(Q)$ gravity strange stars are adiabatically stable for relatively small and positive values of $a$ and vanishing $b$ on the whole interior domain and unstable near the stellar core if we assume that $a\gg0$. On the other hand, adiabatic index shows highly oscillating behavior for non-linear case and there is only small regions of stability for strange stars in this kind of gravity for every negative $a$ (this could be solved in the limit $a\to\infty$, with that option star is adiabatically stable in the intermediate and envelope regions). We plot adiabatic index for both cases on the Figures \ref{fig:7} and \ref{fig:14}.
    \item \textbf{Surface redshift}: it is generally known that surface redshift for physically viable compact stars must not exceed the value of 2, which was respected for both linear and non-linear forms of $f(Q)$. The more detailed graphical representation of the results is illustrated on the first plot of Figures \ref{fig:8} and \ref{fig:15}.
    \item \textbf{Matter causality}: causality conditions need to be respected at the each point of spacetime in order to prevent the appearance of tachyons. From the numerical analysis of linear and non-linear gravities we noticed that strange stars in the aforementioned form of $f(Q)$ does respect causality conditions. Numerical solutions are respectively placed on the second plot of Figures \ref{fig:8} and \ref{fig:15}.
\end{itemize}
Using the aforementioned conclusions on the each quantity of interest (energy conditions, EoS, gradients, matter stability and causality) we could conclude that for linear and non-linear gravities strange star solutions are physically viable and non-singular. However, in the near future it will be of special interest to probe such stellar configurations in the symmetric teleparallel gravity in the presence of additional matter fields, such as inflation or more exotic fields, such as ghost scalar field, QCD Axions and electromagnetic field respecting $U(1)$ local gauge symmetry.

\section*{Data Availability Statement}

There is no new data associated with this article.

\section*{Acknowledgement}
PKS acknowledges National Board for Higher Mathematics (NBHM) under Department of Atomic Energy (DAE), Govt. of India for financial support to carry out the Research project No.: 02011/3/2022 NBHM(R.P.)/R \& D II/2152 Dt.14.02.2022. We are very much grateful to the honorable referees and to the editor for the illuminating suggestions that have significantly improved our work in terms of research quality, and presentation.
\end{document}